\documentclass[aps,prb,superscriptaddress,twocolumn,floatfix]{revtex4-2}
\usepackage{units}
\usepackage{amsmath}
\usepackage{amssymb}
\usepackage{graphicx}
\usepackage{bm}
\usepackage{multirow,color,relsize,ulem,microtype,xr}

\newcommand{\be}{\begin{equation}}
\newcommand{\ee}{\end{equation}}

\newcommand{\re}[1]{\text{Re}[#1]}
\newcommand{\im}[1]{\text{Im}[#1]}

\newcommand{\cc}[1]{{\color{black}#1}}

\begin{document}

\title{Non-Hermitian gauged reciprocity and symmetry}

\author{Jiecheng Lyu}
\affiliation{Department of Materials Science and Engineering, University of Pennsylvania, Philadelphia, PA 19104, USA}

\author{Zihe Gao}
\affiliation{Department of Materials Science and Engineering, University of Pennsylvania, Philadelphia, PA 19104, USA}
\affiliation{Department of Electrical and Computer Engineering, Auburn University, Auburn, AL 36849, USA}

\author{Liang Feng}
\affiliation{Department of Materials Science and Engineering, University of Pennsylvania, Philadelphia, PA 19104, USA}

\author{Li Ge}
\email{li.ge@csi.cuny.edu} 
\affiliation{Department of Physics and Astronomy, College of Staten Island, CUNY, Staten Island, NY 10314, USA}
\affiliation{The Graduate Center, CUNY, New York, NY 10016, USA}

\date{\today}

\begin{abstract} 
The Lorentz reciprocity is a fundamental property in electromagnetism and well known to break down due to an external magnetic field. With a fictitious or imaginary vector potential, however, its behavior is largely unknown. Here we show that in systems with an imaginary vector potential and displaying the non-Hermitian skin effect, the Lorentz reciprocity is broken but still governed by a rigorous mathematical relation, which we term non-Hermitian gauged reciprocity. When mimicking an imaginary vector potential using just linear integrated photonic elements, however, the conditions that lead to the Lorentz reciprocity are still satisfied and hence the latter cannot be broken. Nevertheless, we show that the non-Hermitian gauged reciprocity can still be observed with a proper choice of inputs and outputs, alongside the Lorentz reciprocity. In addition, we also reveal another equal-amplitude response in the same system, which we attribute to a non-Hermitian gauged symmetry. Furthermore, we show that light propagation is not impinged by the non-Hermitian topological funnel effect, highlighting an underappreciated difference between coherently driven and non-driven systems. These findings are confirmed using a tight-binding model and full-wave simulations of coupled optical micro-ring resonators, providing a valuable extension of the Lorentz reciprocity in the non-Hermitian domain. 
\end{abstract}

\maketitle

\section{Introduction}
Wave reciprocity is a familiar concept that has been studied for nearly two centuries \cite{Collin,Landau,Haus,Stumpf}. First discussed by Stokes and Helmholtz for light and by Lord Rayleigh for sound, reciprocity states that the perception of these waves does not depend on the direction of propagation under normal conditions. More commonly known as the Lorentz reciprocity in the context of Maxwell's equations, this property often requires the system to be time-invariant, linear, and with symmetric permittivity and permeability tensors (e.g., non-magnetic). The study of its violation has been revitalized in recent years in the quest for on-chip optical isolators in optical computing and communications \cite{Fan,Feng2,Fan2,Atom,comment}, without using magneto-optical materials that have compatibility issues with current fabrication techniques employed in integrated circuits \cite{magneto}.   

At the same time, non-reciprocity in coupled-mode theory or tight-binding model (such as the Hatano-Nelson model \cite{Hatano,Hatano2}) has been widely used in the study of non-Hermitian physics \cite{NPreview}, leading to the so-called non-Hermitian skin effect \cite{Longhi_gauge,Feng_gauge,Feng_gauge2,Song_gauge}. Featuring a localized mode profile similar to other topological edge states \cite{Hasan,Qi,Alicea,Beenakker,Sarma_RMP}, this non-Hermitian effect is usually demonstrated using direction-dependent couplings and can exist in the absence of a traditional bulk bandgap. As a result, it is a property not of a few midgap states but rather of a significant proportion (if not all) of the system's eigenmodes, which holds in one dimension (1D) and higher dimensions as well, indicating a different mechanism of localization. This mechanism is often referred to as an imaginary gauge transformation: Starting from a corresponding system $\bar{H}$ with reciprocal couplings, this transformation superposes an exponential envelope on top of the eigenstates of $\bar{H}$ in space, which is equivalent to applying an imaginary vector potential. 

Because non-reciprocal couplings in real space are difficult to achieve in atomic, molecular, and condensed matter systems \cite{Gou,Li_gauge,Franca}, the demonstration of non-Hermitian skin effects so far mainly relied on classical wave systems, including mechanical 
\cite{mechanical}, acoustic \cite{acoustic}, and photonic platforms \cite{funnel,Gao}. Therefore, unlike the originally proposed superconducting system in the presence of defects and an external magnetic field by Hatano and Nelson \cite{Hatano}, the apparent breaking of reciprocal couplings between certain modes in such linear wave systems, surprisingly, does \textit{not} violate the conditions that lead to the Lorentz reciprocity. 

Take the aforementioned photonic realization, for example. Non-reciprocal couplings between two whispering-gallery modes of the same chirality are achieved using an auxiliary ring in the middle, and the coupling in one direction is via the upper half of this auxiliary ring while that in the opposite direction is via the bottom half \cite{Longhi_gauge,Feng_gauge,Feng_gauge2}. By imposing gain and loss on these two halves respectively, waves are amplified or attenuated correspondingly when passing through the auxiliary ring in the two opposite directions, leading to different effective coupling strengths. Although optical gain and loss are fundamental in a multitude of nonlinear effects, they do not necessarily invoke nonlinearity or break the Lorentz reciprocity in the linear regime. 

In this work, we show that this discrepancy between non-reciprocal couplings and the persistence of the Lorentz reciprocity is due to the internal degree of freedom of the coupling elements, i.e., the pseudospin associated with clockwise (``spin-down'') and counterclockwise (``spin-up'') modes. More specifically, the Lorentz reciprocity involves both pseudospins, while the non-reciprocal couplings are defined within each pseudospin species. Consequently, while the Lorentz reciprocity still holds in an integrated linear photonic system with non-Hermitian skin modes, the non-reciprocal couplings and effective imaginary magnetic field do lead to a broken reciprocity in transmission in the subspace of one pseudospin, which can be observed with a proper choice of input and output ports. 

More importantly, we show that this broken reciprocity in transmission is still governed by a rigorous mathematical relation, which we term non-Hermitian gauged reciprocity. Furthermore, we show that the transmission of light with a coherent drive is not impinged by the non-Hermitian funnel effect \cite{funnel}, highlighting an underappreciated difference between driven and non-driven non-Hermitian systems. Finally, we reveal an equal-amplitude response different from the Lorentz reciprocity in the same photonic system. We attribute it to a non-Hermitian gauged symmetry, meaning that a system only acquires a symmetry (such as parity) after an imaginary gauge transformation. 

\section{Results} 

\subsection*{Non-Hermitian gauge transformation} 

Since a non-Hermitian gauge transformation is central to our results, we start by briefly reviewing the concept of gauge transformation in electromagnetism and the tight-binding model. Consider the Hamiltonian of a non-relativistic charge particle $q$ in an electromagnetic field:
\be
H = -\frac{\hbar^2}{2m} \left[\vec{\nabla}-\frac{iq}{\hbar c}\vec{A}\right]^2 + qV.
\ee
Here $\vec{A}$ and $V$ are the electromagnetic vector and scalar potential, respectively. The standard gauge transformation in electromagnetism, defined by $\vec{A}\rightarrow\vec{A}'= \vec{A} - \nabla f(\vec{r},t)$ and $V\rightarrow V'= V + \partial_t f(\vec{r},t)$, keeps the electric and magnetic fields invariant. When $ f$ does not depend on time, the scalar potential is then unaffected, and the  difference between $H$ and the system after the gauge transformation, i.e.,
\be
\bar{H} = -\frac{\hbar^2}{2m} \left[\vec{\nabla}-\frac{iq}{\hbar c}\vec{A'}\right]^2 + qV,
\ee
is the phase between their eigenstates, i.e.
\be
\psi_\mu(\vec{r})\rightarrow\bar{\psi}_\mu(\vec{r}) = \psi_\mu(\vec{r})\exp(-iq f(\vec{r})/\hbar c).\label{eq:phase}
\ee
If $ f$ is taken as real (complex), then $\bar{H}$ is Hermitian (non-Hermitian) after the gauge transformation. \cc{In either case, the energy eigenvalues remain unchanged, even though the probability amplitude is changed when the gauge field $f$ is complex. This change leads to physically observable outcomes, and it is at the core of the non-Hermitian skin effect, where the probability amplitude of any eigenstate in the system is pushed towards one edge or corner determined by the gauge field.} 

A similar gauge transformation can be defined for the following tight-binding model of a one-dimensional chain with the open boundary condition:
\be
H = \sum_{j} V_j|j\rangle\langle j| \,+\, \bar{t}_{j,j+1}\hspace{-3pt}\left(e^{i\alpha_j}|j+1\rangle\langle j| + e^{-i\alpha_j}|j\rangle\langle j+1|\right), \label{eq:H}
\ee
$V_j$, the onsite potential, plays the role of the scalar potential in electromagnetism, while $\alpha_j$ acts as the vector potential, which is evaluated at the middle between sites $j$ and $j+1$. Its gauge transformation can be defined similarly to that in electromagnetism, i.e., $\alpha'_j=\alpha_j-(\nabla f)_j$, and it is easy to show that 
\be
\bar{H} = \sum_{j=1}^{N} V_j|j\rangle\langle j| \,+\, \bar{t}_{j,j+1}\hspace{-3pt}\left(e^{i\alpha'_j}|j+1\rangle\langle j| + e^{-i\alpha'_j}|j\rangle\langle j+1|\right) \nonumber
\ee
has the same energy eigenvalues as $H$. 

This gauge transformation can be written explicitly as $\bar{H} = \bar{G}^{-1} H \bar{G}$, where $\bar{G}$ is a diagonal matrix with elements $g_{11}=1$ and $g_{jj}=e^{if_{j}}\,(j>1)$. We then find  
\be
g_{jj}\bar{\psi}_\mu(j) =  \psi_\mu(j),\label{eq:amp1}
\ee
where $\bar{\psi}_\mu(j)$, $\psi_\mu(j)$ are the wave functions of $\bar{H}$ and $H$ at site $j$. Note that we have used $(\nabla f)_j \equiv (f_{j+1}- f_{j})/\Lambda$ and set the lattice constant $\Lambda=1$, or equivalently, $f_j=\sum_{n=1}^{j-1}(\nabla f)_n$ with the boundary condition $f_1=0$. 

Let us come back to the tight-binding Hamiltonian $H$ in Eq.~(\ref{eq:H}) before the gauge transformation. We denote its nearest-neighbor (NN) couplings as
 $t_{j+1,j} = \bar{t}_{j,j+1} e^{i\alpha_j},\,t_{j,j+1} = \bar{t}_{j,j+1} e^{-i\alpha_j}$. 
\cc{In other words, $\bar{t}_{j,j+1}$ is defined as the geometric average of $t_{j+1,j}$ and $t_{j,j+1}$}. The vector potential can then be expressed as
\be
\alpha_j = \frac{i}{2}\log\frac{t_{j,j+1}}{t_{j+1,j}}, \label{eq:A}
\ee
and when the couplings $t_{j,j+1}$,$t_{j+1,j}$ are non-reciprocal and differ only by amplitude, the vector potential $\alpha_j$ is then imaginary. Such a system is hence often referred to as an imaginary or non-Hermitian gauged array \cite{Gao,Ge_PRB2023}.

\subsection*{Coherent drive and wave propagation}

To quantify wave transmission in the presence of an imaginary magnetic field, we discuss below wave propagation excited by a coherent drive in a tight-binding model:
\begin{align}
\hspace{-2mm} i\dot{a}_1 &= (\omega_1 - \omega) a_1 + t_{12} a_2 - i\sqrt{\kappa'} a_1^{(i)} \nonumber\\
\hspace{-2mm} i\dot{a}_2 &= (\omega_2 - \omega) a_2 + t_{21} a_1 + t_{23} a_3  \label{eq:CMT}\\
&\ldots \nonumber\\
\hspace{-2mm} i\dot{a}_N &= (\omega_N - \omega) a_{N} + t_{N,N-1} a_{N-1} - i\sqrt{\kappa'} a_N^{(i)}\nonumber.
\end{align}
Here $a_{j}$ $(j=1,2,\ldots)$ is the wave amplitude at site $j$ in the rotating frame of the drive, which has frequency $\omega$. The overhead dots denote the time derivative, and $\omega_j$ is a complex resonance at site $j$, including both the real-valued resonant frequency and the intrinsic loss rate. 

Compared with the tight-binding model in Eq.~(\ref{eq:H}), the set of equations above have two additional terms proportional to $\sqrt{\kappa'}$, which represent the drives at the left and right of the system: When we drive this 1D chain from the left (right), we set $a_1^{(i)}=1$, \cc{$a_N^{(i)}=0$} ($a_1^{(i)}=0$, \cc{$a_N^{(i)}=1$}) and denote the output at the opposite end by $a_N^{(\text{o})}$ ($a_1^{(\text{o})}$). The latter is given by $a_N^{(\text{o})} = \sqrt{\kappa'}a_N$ and $a_1^{(\text{o})} = \sqrt{\kappa'}a_1$, where $\kappa'$ is the coupling loss to the input and output channels (e.g., waveguides) and $a_N$, $a_1$ are again the wave amplitudes at the last and first site. 

\cc{When we consider reciprocity in this framework, the time-reversal of one input (output) channel is taken as the new output (input) channel in the opposite direction. This convention is used extensively in the study of the scattering matrix \cite{scatteringM} in various systems.}

Below we consider $\omega_j$'s close in frequency (if not identical), and they couple to form $N$ supermodes, where $t_{ij}$ is the coupling from site $j$ to $i$ and different in magnitude from $t_{ji}$ in the opposite direction. The coherent drive excites one or more of these supermodes, and the steady state of Eq.~(\ref{eq:CMT}) is found by setting all $\dot{a}_j=0$. It can be conveniently solved using 
\be
(H - \omega \bm{1}) \bm{a} = i\sqrt{\kappa'} \bm{a}^{(i)} \label{eq:CMT_a}
\ee
where $\bm{a} = [a_1,a_2,\ldots,a_N]^T$, $\bm{a}^{(i)} = [a_1^{(i)}, 0, \ldots, 0,a_N^{(i)}]^T$. Here $H$ is the same Hamiltonian given by Eq.~(\ref{eq:H}). 

Different from the standard technique used in temporal coupled-mode theory \cite{TCMT} that expresses $\bm{a}$ by inverting the matrix operator on the left hand side of Eq.~(\ref{eq:CMT_a}), we use the left eigenstate $\tilde{\psi}_\mu^T$'s and right eigenstates $\psi_\mu$'s of $H$ instead: 
\be
\bm{a} = \sum_\mu \frac{\sqrt{\kappa'}}{\lambda_\mu-\omega}[\tilde{\psi}_\mu^T\bm{a}^{(i)}]\psi_\mu.\label{eq:ss}
\ee  
Here $\tilde{\psi}_\mu$ and $\psi_\mu$ are defined by \cite{Bronson}
\be
H^T\tilde{\psi}_\mu = \lambda_\mu \tilde{\psi}_\mu,\quad H{\psi}_\mu = \lambda_\mu {\psi}_\mu,\label{eq:LeftEig}
\ee
where the superscript $``T"$ denotes the matrix transpose. We have assumed that the system is away from an exceptional point \cite{NPreview}, leading to the biorthogonal relation $\tilde{\psi}_\mu^T{\psi}_\nu=\delta_{\mu\nu}$. It is through the properties of the left and right eigenstates of $H$ that we reveal the unique non-Hermitian properties of wave propagation in our system in the following sections. 

The transmission coefficient in this system from left to right and that in the opposite direction can be defined by
\be
t_L\equiv\frac{a^{(\text{o})}_N}{a^{(i)}_1} = \frac{\sqrt{\kappa'}\bm{a}^{(L)}_N}{1},\quad t_R\equiv\frac{a^{(\text{o})}_1}{a^{(i)}_N} = \frac{\sqrt{\kappa'}\bm{a}^{(R)}_1}{1},\label{eq:ts}
\ee
where $\bm{a}^{(L)}$, $\bm{a}^{(R)}$ are the amplitudes in the steady state when driving the system from the left and right, respectively. 

Below we show explicitly that although the Lorentz reciprocity is broken in this tight-binding model, i.e., $t_L\neq t_R$, it evolves to a non-Hermitian gauged reciprocity. 

\subsection*{Non-Hermitian gauged reciprocity}

For a non-Hermitian gauged array $H$ with $t_{j,j+1}\neq t_{j+1,j}\in\mathbb{R}$ and an imaginary vector potential $\alpha_j$, it is clear from our discussion of Eq.~(\ref{eq:H}) that $H$ can be transformed to a system $\bar{H}$ with reciprocal couplings \cc{$\bar{t}_{j,j+1}=\sqrt{{t}_{j+1,j}{t}_{j,j+1}}\in\mathbb{R}$} and $\alpha'_j=0$. The required function $f$ is also imaginary, given by $f_j=\sum_{n=1}^{j-1}\alpha_n$. Therefore, it is customary to refer to $H\rightarrow\bar{H}=\bar{G}^{-1} H \bar{G}$ as an imaginary gauge transformation, where $\bar{G}$ is the diagonal matrix introduced in Eq.~(\ref{eq:amp1}). 

We also note that unlike in the original Hatano-Nelson model \cite{Hatano}, we do not require $\alpha_j$'s (and $\bar{t}_{j,j+1}$'s) to be identical. More importantly, we note that $H^T$ describes a system that differs from $H$ by only exchanging $t_{j,j+1}$ and $t_{j+1,j}$. As a result, it is easy to show that $H^T$ and $\bar{H}$ are related by the same imaginary gauge transformation, i.e., ${H}^T = \bar{G}^{-1}\bar{H}\bar{G}$. 

Combined with the definition of the left eigenstate $\tilde{\psi}_\mu^T$ mentioned in Eq.~(\ref{eq:LeftEig}), we find 
\be
\bar{G}_{jj}\tilde{\psi}_\mu(j) = \bar{\psi}_\mu(j).\label{eq:Gjj}
\ee
\cc{Note that we have used the site position as the argument of the eigenstates and reserve the subscript for the mode index, which is consistent with the continuous case [see Eq.~(\ref{eq:phase})] and different from our notation for the vector potential $\bm{\alpha}$, the steady state solutions $\bm{a}^{(L)}$,$\bm{a}^{(R)}$, and input vector $\bm{a}^{(i)}$}. 

Inserting Eq.~(\ref{eq:Gjj}) in the steady state given by Eq.~(\ref{eq:ss}), we derive
\begin{align}
\bm{a}^{(L)}_N
&= \sum_\mu \frac{\sqrt{\kappa'}\bar{\psi}_\mu(1)\bar{\psi}_\mu(N)}{\lambda_\mu-\omega}\frac{g_\text{\begin{tiny}NN\end{tiny}}}{g_{11}},  \label{eq:aL}\\
\bm{a}^{(R)}_1 
&= \sum_\mu \frac{\sqrt{\kappa'}\bar{\psi}_\mu(N)\bar{\psi}_\mu(1)}{\lambda_\mu-\omega}\frac{g_{11}}{g_\text{\begin{tiny}NN\end{tiny}}}, \label{eq:aR}
\end{align}
and in turn,
\be
\frac{t_R}{t_L} = g_\text{\begin{tiny}NN\end{tiny}}^{-2} =\prod_{j=1}^{N-1} e^{-2\alpha_j}= \prod_{j=1}^{N-1}\frac{ t_{j,j+1} }{ t_{j+1,j}}, \label{eq:gauged_reciprocity}
\ee
where we have used $g_{11}=1$. This result shows explicitly that the Lorentz reciprocity is broken in the tight-binding model with different NN couplings $t_{j,j+1}\neq t_{j+1,j}\in\mathbb{R}$, and we refer to this relation as non-Hermitian gauged reciprocity. 

We note that this non-Hermitian gauged reciprocity depends only on the imaginary gauge field $\alpha_j$'s, and it is independent of both the underlying system with reciprocal couplings (i.e., $\bar{H}$) \cc{and the driving frequency, even though the transmission coefficients $t_{L,R}$ are [through $\bm{a}^{(L)}_N$ and $\bm{a}^{(R)}_1$ in Eqs.~(\ref{eq:aL}) and (\ref{eq:aR})].} 

\begin{figure}[t]
\centering
\includegraphics[clip,width=\linewidth]{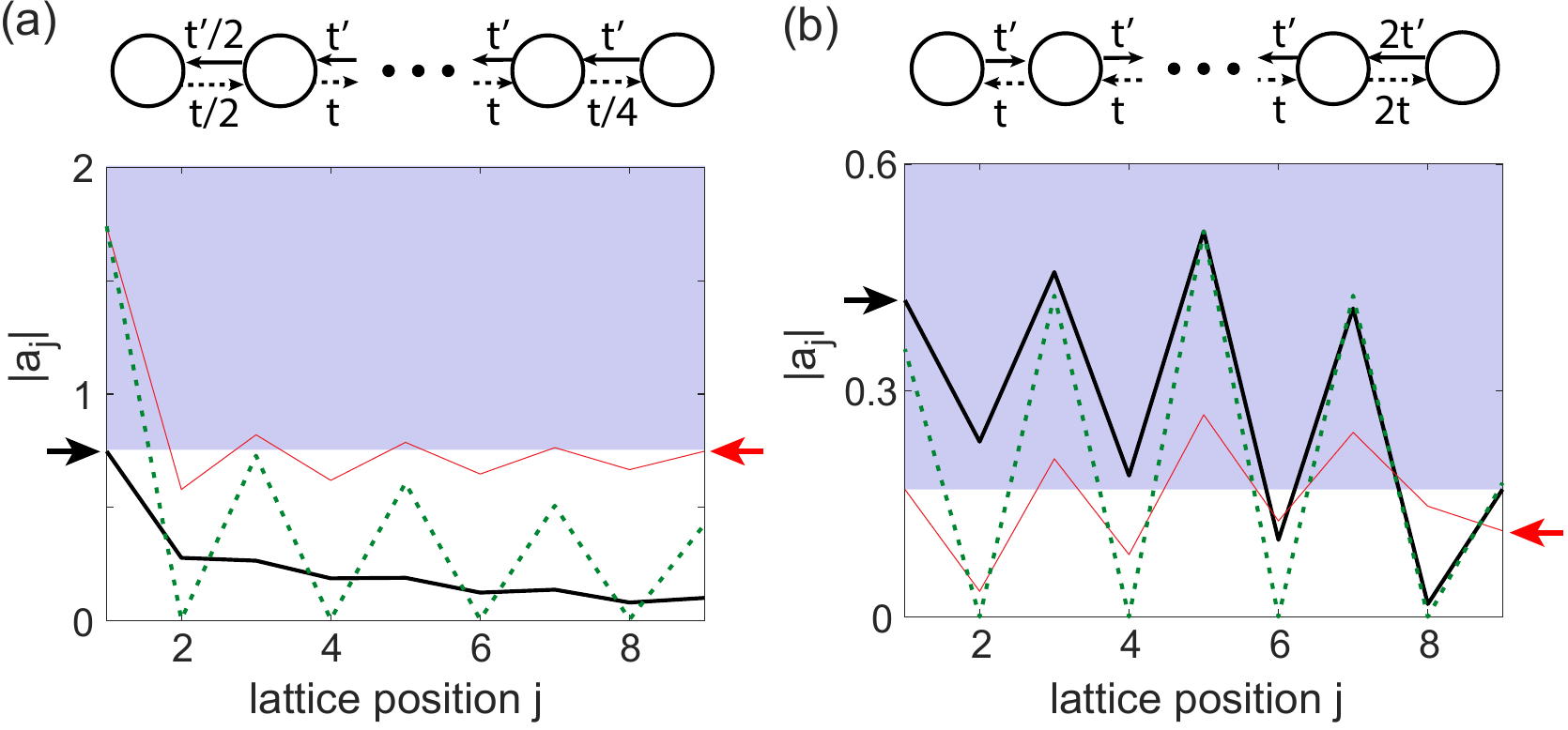}
\caption{\textbf{Broken and survived Lorentz reciprocity in non-Hermitian gauged arrays}. (a) Mode profiles of the steady state when driving the left side (thick solid line; indicated by the arrow on the left) and the right side (thin solid line; indicated by the arrow on the right), respectively. The bottom edge of the shaded area highlights the equal-amplitude response in all figures, either due to non-Hermitian gauged symmetry or Lorentz reciprocity. Dashed line shows the non-Hermitian skin mode at the driving frequency. $t'=1.2t$ and $\kappa_0 = -2\im{\omega_0}=0.4t = 4\kappa'$. The input and output channels are not shown. 
(b) Same as (a) but with a different configuration of non-reciprocal couplings.} \label{fig:gauged}
\end{figure}

The Lorentz reciprocity in the tight-binding model is restored when the product of all couplings to the left equals the product of all couplings to the right, with reciprocal couplings being the trivial case. We illustrate broken and survived Lorentz reciprocity numerically in Fig.~\ref{fig:gauged}, using an array with nine sites ($N=9$) and no detuning (i.e., all $\omega_j$'s equal $\omega_0$), driven at $\omega=\re{\omega_0}$. 
Note that the black and red arrows next to the main figures only indicate the direction of the drive but not its amplitude; the latter is set to 1 in all cases as specified in Eq.~(\ref{eq:ts}).
In the case shown in Fig.~\ref{fig:gauged}(a), the zero mode (i.e., with $\re{\lambda_\mu}=\omega_0$) is localized on the left, \cc{and we can readily see the directional gain and loss due to this non-Hermitian skin effect: the wave amplifies (attenuates) when propagating from right to left (left to right).} Consequently, we find ${t_R}/{t_L}=4\times1.2^8$ 
as Eq.~(\ref{eq:gauged_reciprocity}) predicts. 
\cc{This ratio does not change when we drive at a different frequency, e.g., at the next resonance $\omega/t=0.375$ above the zero mode.}
In the case shown in Fig.~\ref{fig:gauged}(b), indeed we find $t_R=t_L$ by letting the couplings to the right (left) be stronger in the left (right) half of the array, leading to a zero mode peaked at the center \cite{Gao,Ge_gauge}.

\begin{figure}[t]
\centering
\includegraphics[clip,width=\linewidth]{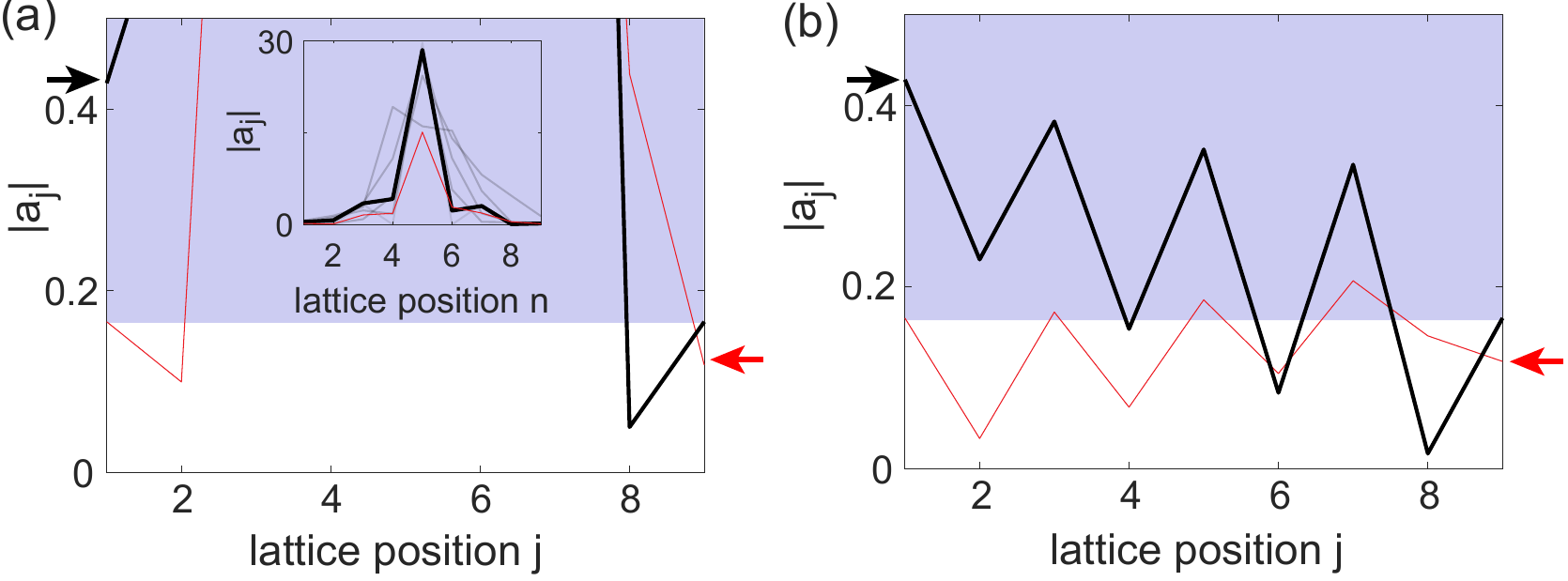}
\caption{\textbf{Identical transmission with and without a non-Hermitian funnel}. (a) Same as Fig.~\ref{fig:gauged}(b) but with $t'=9t$. $\bar{t}=\sqrt{tt'}$ remains the same. Insert: Zoomed-out view showing the whole steady states and all eigenstates (grey lines). (b) Same as (a) but with $t'=t$.} \label{fig:funnel}
\end{figure}

If we increase the contrast between each pair of non-reciprocal couplings in this latter case, all modes of the system peak strongly at the center of the system [Fig.~\ref{fig:funnel}(a); insert], which is a variant of the non-Hermitian skin effect. As a result, any initial excitation localized away from the center will eventually move to the middle, which was termed a non-Hermitian topological funnel \cite{funnel}. Being topological, this behavior seems to imply a universal behavior, whether the system is coherently driven or not. 
We find, instead, that not only can light escape the funnel and propagate from one end of the lattice to the opposite side, in the steady state with a coherent drive as in our consideration above; the transmission coefficients are also independent of the funnel here.

More specifically, these transmission coefficients are identical to the case without the non-Hermitian skin effect (see Fig.~\ref{fig:funnel}), i.e., with each pair of couplings being reciprocal and all modes being extensive across the lattice. This behavior can be understood by first noticing that the factor $g_\text{\begin{tiny}NN\end{tiny}}$ in Eq.~(\ref{eq:gauged_reciprocity}) becomes 1 and hence equal to $g_{11}$. As a result, we find
\be
t_L = \sum_\mu \frac{\kappa'\bar{\psi}_\mu(N)\bar{\psi}_\mu(1)}{\lambda_\mu-\omega} = t_R
\ee    
using Eqs.~(\ref{eq:ts}), (\ref{eq:aL}), and (\ref{eq:aR}). Therefore, these transmission coefficients only depend on the eigenvalues $\lambda_\mu$'s and eigenstates $\bar{\psi}_\mu$'s of the Hamiltonian $\bar{H}$ with reciprocal couplings, i.e., they are independent of the imaginary vector potential and the non-Hermitian funnel at the middle. This observation highlights the contrasting behaviors of driven and non-driven non-Hermitian systems, a difference underappreciated in previous studies.

\cc{We note that we have taken all on-site resonances $\omega_j$'s to be the same (and set them to zero) here, which is common in photonic devices to enhance the coupling of neighboring resonators that is due to resonant tunneling. The non-Hermitian gauged reciprocity, however, does not require this condition, as can be seen from the derivation of Eq.~(\ref{eq:gauged_reciprocity}).}

\subsection*{Non-Hermitian gauged symmetries} 

Despite the lack of the Lorentz reciprocity, the case shown in Fig.~\ref{fig:gauged}(a) clearly displays another equal-amplitude response: the light amplitude in the steady state at the driven site is the same whether we send in light from the left or right, i.e.,
\be
\bm{a}^{(L)}_1 = \bm{a}^{(R)}_N. \label{eq:Theorem2}
\ee
To identify the symmetry leading to this phenomenon, we note $\tilde{\psi}_\mu(j) \psi_\mu(j) = \bar{\psi}^2_\mu(j)$ and 
\be
\bm{a}^{(L)}_1 
= \sum_\mu \frac{\sqrt{\kappa'}}{\lambda_\mu-\omega} \bar{\psi}^2_\mu(1), \quad
\bm{a}^{(R)}_N 
= \sum_\mu \frac{\sqrt{\kappa'}}{\lambda_\mu-\omega} \bar{\psi}^2_\mu(N), \nonumber
\ee
both expressed using the wave function $\bar{\psi}_\mu$ of the \textit{reciprocal} system $\bar{H}$ only. Equation~(\ref{eq:Theorem2}) holds when  
\be
\bar{\psi}_\mu(1) = \pm\bar{\psi}_\mu(N) \label{eq:Theorem2_cond}
\ee
is true for \textit{all} eigenstates of $\bar{H}$, which demands that $\bar{H}$ is parity symmetric. 

In other words, the equal-amplitude response given by Eq.~(\ref{eq:Theorem2}) requires the system $H$ to have a \textit{non-Hermitian gauged symmetry}: it does not have parity symmetry due to its non-reciprocal couplings, but it does so after the non-Hermitian gauge transformation $\bar{G}$:
\be
P^{-1}\bar{H}P = \bar{H},\quad\bar{H}=\bar{G}^{-1}H\bar{G}.\label{eq:gaugedP0}
\ee
\cc{For example, all (reciprocal) couplings in $\bar{H}$ corresponding to the model used in Fig.~\ref{fig:gauged}(a) are given by $\sqrt{tt'}$, except for the leftmost and rightmost one, both given by $\sqrt{tt'}/2$}. As a result of this non-Hermitian gauge transformation, \cc{the probability amplitude in an eigenstate is changed, which is evident from the requirement imposed by Eq.~(\ref{eq:Theorem2_cond}), but the corresponding energy eigenvalue remains the same as we mentioned before.} 

We can also express the symmetry relation (\ref{eq:gaugedP0}) as
\be
\bar{P}^{-1}H\bar{P} = H, \quad \bar{P} \equiv \bar{G}P\bar{G}^{-1},\label{eq:gaugedP}
\ee
and we refer to $\bar{P}$ as a non-Hermitian gauged parity symmetry. \cc{This proof shows that the equal-amplitude response due to the non-Hermitian gauged parity symmetry holds at any driving frequency, not just at the resonance of the zero mode;} it also permits on-site detuning of $\omega_j$'s, as long as they remain parity symmetric, which are unaffected by the non-Hermitian gauge transformation.

\begin{figure*}[t]
\centering
\includegraphics[width=\linewidth]{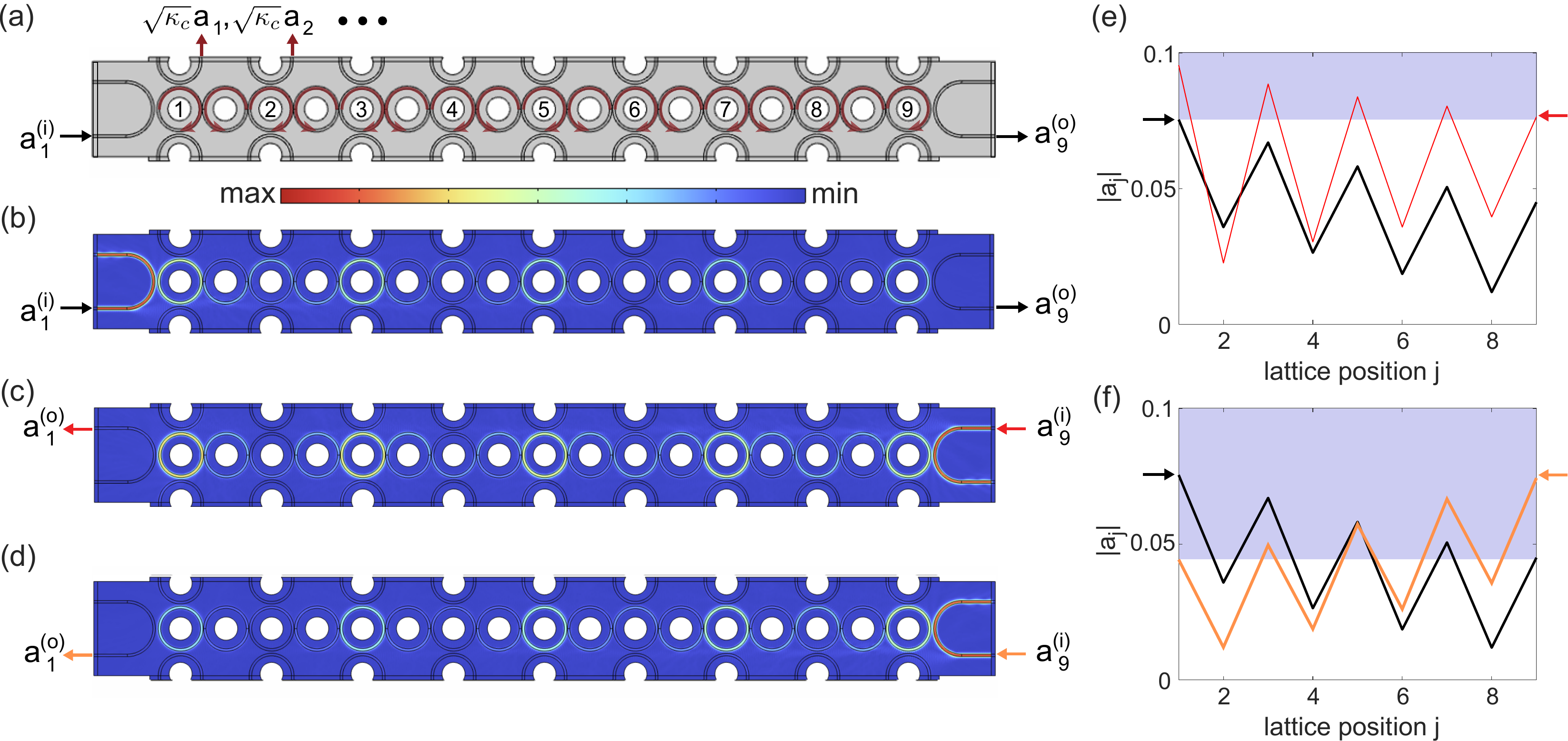}
\caption{\textbf{Photonic simulations of a coherently driven non-Hermitian gauged array}. (a) Schematic of the coupled micro-rings and input/output channels. See Sec.~\ref{sec:zeromode} for the parameters used. (b-d) Field patterns when driving from the bottom left, top right, and bottom right waveguide, respectively. $|E|$ is shown. (e) Verification of non-Hermitian gauged reciprocity [Eq.~(\ref{eq:gauged_reciprocity})] and symmetry [Eq.~(\ref{eq:Theorem2})] given by the tight-binding model. (f) Verification of the Lorentz reciprocity in the actual device. Amplitudes in (e,f) are normalized by the input and from the stronger half of each ring.}
\label{fig:gauged_simulation}
\end{figure*}

The concept of non-Hermitian gauged symmetries can be easily generalized by replacing the parity in Eq.~(\ref{eq:gaugedP}) with another symmetry, whether it is a linear symmetry such as rotation or an anti-linear symmetry such as parity-time ($PT$) symmetry. For example, a non-Hermitian chiral symmetry $\Pi$ can be constructed using the product of $PT$ symmetry and non-Hermitian particle-hole symmetry (NHPH) $CT$ \cite{zeromodeLaser,Defect}, i.e., $\Pi = (PT)(CT) = { PC}$. $PT$ is broken if we apply a non-Hermitian gauge transformation and turn reciprocal couplings into non-reciprocal couplings, but the system now has non-Hermitian gauged $PT$ symmetry instead, i.e., $\bar{G}{PT}\bar{G}^{-1} = \bar{ P} { T}$. At the same time, reciprocal couplings are not required by NHPH symmetry, and hence the latter remains a symmetry in the presence of the non-reciprocal couplings. As a result, the system's chiral symmetry evolves to $\bar{\Pi} = (\bar{ P} { T})({CT}) = \bar{G} { P}\bar{G}^{-1}{ C} = \bar{G}\Pi\bar{G}^{-1}$, which is also non-Hermitian gauged.

\section{Photonic simulations} 

As we mentioned in the introduction, the realization of the tight-binding model in Fig.~\ref{fig:gauged}(a) with non-reciprocal couplings on a linear photonic platform \textit{cannot} break the Lorentz reciprocity. 

To understand the discrepancy between this general property and the non-Hermitian gauged reciprocity in the tight-binding model with non-reciprocal couplings, here we examine the implications of the tight-binding model. This model considers the couplings of a set of \textit{fixed} modes as we have seen in Eq.~(\ref{eq:CMT}), independent of the direction of wave propagation. As an example, consider the case shown in Fig.~\ref{fig:gauged_simulation}(a), \cc{where the input $a_1^{(i)}$ from the U-shaped waveguide on the left drives a clockwise (CW) mode in the first micro-ring resonator}. This array has $9$ ``cavity'' rings [numbered in Fig.~\ref{fig:gauged_simulation}(a)] captured by the $9$ sites in the tight-binding model, and the non-reciprocal couplings between their CW modes are realized by coupling to the counterclockwise (CCW) modes in the $8$ auxiliary rings [unnumbered in Fig.~\ref{fig:gauged_simulation}(a)] as mentioned in the introduction. The output $a_9^{(\text{o})}$ on the right in the tight-binding model is then from the bottom of the right waveguide. Now to compare with the result obtained from driving from the right side in the tight-binding model, we need to excite the \textit{same} set of CW modes in the cavity rings. The input light then needs to be sent in from the top of the right waveguide [Fig.~\ref{fig:gauged_simulation}(c)], which is \textit{not} the output channel when we drive from the left. 

This observation highlights that the discrepancy between non-reciprocal couplings in the tight-binding model and the persistence of the Lorentz reciprocity on a linear photonic platform is due to the internal degree of freedom of the coupling elements, i.e., the pseudospin associated with CW (spin-down) and CCW (spin-up) modes. 

To demonstrate both the non-Hermitian gauged reciprocity in the subspace of one pseudospin species and the persisting Lorentz reciprocity, we simulate wave propagation in this photonic system using the finite-element method (COMSOL Multiphysics). Semi-circular waveguides are added to the top and bottom sides of the array to measure the amplitude in each ring [e.g., those marked by $\sqrt{\kappa_c}a_{1,2}$ in Fig.~\ref{fig:gauged_simulation}(a)], and the additional loss $\sqrt{\kappa_c}$ due to this scheme can be included in the total loss rate of each ring, i.e., $\im{\omega_j}$ in Eq.~(\ref{eq:CMT}). The input/output waveguides on the left and right are also curved to ensure that they do not couple directly to the semi-circular waveguides. 

Comparing the two driving scenarios consistent with the tight-binding model [Figs.~\ref{fig:gauged_simulation}(b,c)], we indeed observe the equal-amplitude response warranted by non-Hermitian gauged parity symmetry [Fig.~\ref{fig:gauged_simulation}(e)], \cc{together with the directional gain (loss) when light propagates from right to left (left to right)}. To verify the non-Hermitian gauged reciprocity given by Eq.~(\ref{eq:gauged_reciprocity}), we note that instead of characterizing each pair of non-reciprocal couplings, $g_\text{\begin{tiny}NN\end{tiny}}$ can be obtained from the ratio of the amplitudes in the first and last ring in the zero mode (see Sec.~\ref{sec:zeromode}), which we found to be approximately $1.48$. Its square (i.e., $2.19$) agrees well with $t_R/t_L=\bm{a}^{(R)}_1/\bm{a}^{(L)}_9\approx2.12$ extracted from Fig.~\ref{fig:gauged_simulation}(e). This result also confirms directly the breaking of the Lorentz reciprocity in the tight-binding model. 

To verify the standard Lorentz reciprocity of this device, however, we need to send in light from the bottom of the right waveguide [Fig.~\ref{fig:gauged_simulation}(d)], and it excites the CCW modes in the cavity rings instead, which are coupled by the CW modes in the auxiliary rings. \cc{Since the directions of wave propagation are reversed in the two halves of the auxiliary rings, each pair of non-reciprocal couplings are also switched in the tight-binding model. In other words, flipping the pseudospin leads to the flip of the sign of the vector potential $\alpha_j$'s in Eq.~(\ref{eq:H}) as well.

Once one realizes this point, it is easy to show via Eqs.~(\ref{eq:aL}) and (\ref{eq:aR}) that the two transmission coefficients $t_L$ and $t_R$ also exchange when we flip the pseudospin, i.e., $t_L$ ($t_R$) for the spin-down modes becomes $t_R$ ($t_L$) for the spin-up modes, which proves the Lorentz reciprocity [see Fig.~\ref{fig:gauged_simulation}(f)]. 

Note that this derivation does not require the Hamiltonian for one pseudospin to be the mirror image of that for the other pseudospin, as dictated by the generality of the Lorentz reciprocity. In other words, their shared $\bar{H}$ with reciprocal couplings does not need to have parity, or equivalently, themselves are not required to have non-Hermitian gauged parity symmetry.}

One intriguing feature of the field patterns in the photonic simulations shown in Fig.~\ref{fig:gauged_simulation} is that a subset of cavity rings, despite having a uniform refractive index, display contrasting amplitudes between the top and bottom halves, which cannot be captured by a single eigenmode of the ring resonator, an assumption used in the tight-binding model. Nevertheless, the observed non-Hermitian gauged reciprocity and symmetry do not rely on this assumption, and they can be derived analytically on this photonic platform using a transfer matrix approach (see Sec.~\ref{sec:Tmatrix}). 

\section{Conclusion} 

In summary, we have discussed the transmission properties of non-Hermitian gauged arrays when responding to a coherent drive, both in the usual tight-binding model and a prominent integrated photonic platform with coupled micro-ring resonators. While the Lorentz reciprocity is broken in the former in general, it is replaced by a non-Hermitian gauged reciprocity that depends only on the imaginary vector potential. This breaking of the Lorentz reciprocity and the emergence of its replacement require driving modes of the same pseudospin when the wave propagates in the two opposite directions. The actual Lorentz reciprocity, on the other hand, involves one pseudospin when light propagates from left to right and the other pseudospin in the opposite propagation direction. As a result, both types of reciprocity can be observed in the same photonic system, as we have shown using full-wave simulations. 

\cc{Here we also note that the general principle of reciprocity does not require time reversal, a related property that is absent in most photonic systems. For example, if we replace non-reciprocal couplings in Fig.~\ref{fig:gauged} by reciprocal couplings, then Eq.~(\ref{eq:gauged_reciprocity}) tells us that $t_R/t_L=1$, i.e., the transmission is reciprocal. This result holds independent of whether the system is lossy. If it is, e.g., with a negative imaginary part for each on-site frequency $\omega_j$, then time-reversal symmetry of the system is lifted; its time-reversal partner has optical gain at each site instead, i.e., with a positive imaginary part for each $\omega_j$.} 

\cc{By the same argument, the independence of time reversal (i.e., $\im{\omega_j}$'s) should also hold for the non-Hermitian gauged reciprocity, but one is reminded that non-reciprocal couplings indicate intrinsic energy exchange with the environment at the coupling junctions \cite{Ge_PRB2023}, and hence time reversal symmetry is lifted by default. This energy exchange at the coupling junctions is also crucial to understand the non-Hermitian skin effect, and in particular, the directional gain (and loss) we have seen in both the tight-binding model [Fig.~\ref{fig:gauged}(a)] and the photonic simulation [Fig.~\ref{fig:gauged_simulation}(e)]: The loss at each site cannot possibly produce gain as the wave propagates, and therefore, it must come from the coupling junctions.}

Furthermore, by contrasting with the non-Hermitian funnel effect, we have highlighted the difference between driven and non-driven non-Hermitian systems that was underappreciated in previous studies. \cc{This difference is not due to the excitation of multiple modes in the steady state: One can choose to excite any mode preferably by setting the drive frequency at the resonant frequency of that mode. When this mode is well separated from the other modes spectrally and has a low loss, the steady state with the drive is essentially given by just this mode alone as can be seen from the fraction in Eq.~(\ref{eq:ss}). If this mode has the lowest loss (or highest gain), it then matches the surviving mode in the long run of the non-driven case that manifests the non-Hermitian funnel effect, both localized at the position of the funnel [see Fig.~\ref{fig:funnel}(a); inset]. 

The reason that light can propagate through the funnel and reach the other side in the steady state, instead, is just the presence of the drive, which sends light consistently into the system despite its buildup at the funnel position. The observation that the transmission coefficients are independent of the funnel, i.e., they are the same in both $H$ with the funnel and $\bar{H}$ without the funnel, is a neat consequence of the imaginary gauge transformation between $H$ and $\bar{H}$ where $g_{11}=g_\text{\begin{tiny}NN\end{tiny}}=1$.} 

Finally, we have also revealed a non-Hermitian gauged symmetry in the same system, which leads to a different equal-amplitude response to the coherent drive. These observations provide a valuable perspective on the roles of an imaginary vector potential on the non-Hermitian extension of the Lorentz reciprocity, which may help in designing future functional photonic devices. 
\\

\begin{acknowledgements}
L.G. acknowledges support by the National Science Foundation (NSF) under Grants No. PHY-1847240 and No. DMR-2326698. J.L., Z.G., and L.F. acknowledge support by NSF under Grant No. DMR-2326699.
\end{acknowledgements}

\appendix

\section{Non-Hermitian zero mode}
\label{sec:zeromode}

\begin{figure*}[t]
\includegraphics[clip,width=\linewidth]{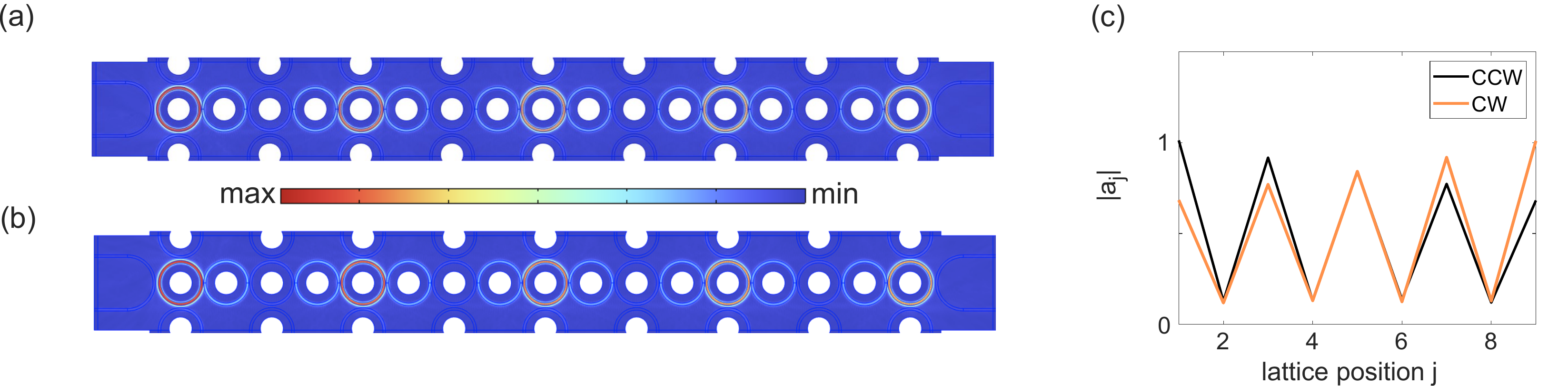}
\caption{\textbf{Spatial profiles of the zero modes in Fig.~\ref{fig:gauged_simulation} of the main text}. (a,b) Field patterns of the CW and CCW zero modes, showing their $|E|$. (c) Extracted CW and CCW amplitudes in the two zero modes, showing $g_{9,9}\approx1.48$.
} \label{fig:zeromode}
\end{figure*}

Due to the non-Hermitian particle-hole (NHPH) symmetry mentioned in the main text and the Lieb theorem \cite{LiebTheorem}, the system shown in Fig.~\ref{fig:gauged}(a) of the main text has a zero mode, i.e., with $\lambda_\mu=\omega_0$, and so does the corresponding photonic system shown in Fig.~\ref{fig:gauged_simulation}. We note that the imaginary part of $\omega_0$, i.e., $-\kappa_0/2$, already includes the waveguide coupling loss in the first and the last rings in the tight-binding model, and for our integrated photonic platform, $\kappa_0$ also includes the coupling loss to the semi-circle waveguides. 

This zero mode $\psi_\mu$ and its partner $\bar{\psi}_\mu$ in the corresponding Hamiltonian $\bar{H}$ with reciprocal coupling $\bar{t}$ mentioned in the main text are related by:
\begin{gather}
\psi_\mu(1) = g_{11}\bar{\psi}_\mu(1) = \bar{\psi}_\mu(1),\\
\psi_\mu(N) = g_\text{\begin{tiny}NN\end{tiny}}\bar{\psi}_\mu(N)=g_\text{\begin{tiny}NN\end{tiny}}\bar{\psi}_\mu(1).
\end{gather}
In the last step, we have used $\bar{\psi}_\mu(1)=\bar{\psi}_\mu(N)$ because $\bar{H}$ has parity symmetry and this mode is an even parity mode when $N=9$. Therefore, instead of obtaining $g_\text{\begin{tiny}NN\end{tiny}}$ via the product in Eq.~(\ref{eq:gauged_reciprocity}) of the main text by characterizing  every pair of non-reciprocal couplings on the photonic platform, a much easier way is to derive it from the ratio of $\psi_\mu(N)/\psi_\mu(1)$ in this zero mode [see Fig.~\ref{fig:zeromode}(c)]. 

This zero mode we consider here consists of CW (``spin-down'') modes in the cavity rings, which we show in Fig.~\ref{fig:zeromode}(a). It also has a degenerate partner consisting of CCW (``spin-up'') modes in the cavity rings [Fig.~\ref{fig:zeromode}(b)], which is excited in Fig.~\ref{fig:gauged_simulation}(d) in the main text when verifying the Lorentz reciprocity of the photonic platform by sending in light from the bottom of the right waveguide. 

The parameters used in the photonic system shown in Fig.~\ref{fig:gauged_simulation} in the main text and here in Fig.~\ref{fig:zeromode} are the same. The outer radius and refractive index of the cavity and auxiliary rings are 4 $\mu$m, $2.8$, and 4.038 $\mu$m, $2.8\pm0.0012i$, respectively. The ring spacing is 0.08  $\mu$m, waveguide-ring spacing is 0.5 $\mu$m, and the driving wavelength is $1515.35$ nm. The background refractive index is 1.9, and the width of all rings and waveguides is 0.45 $\mu$m.

\section{Transfer matrix analysis}
\label{sec:Tmatrix}

\begin{figure}[b]
\includegraphics[clip,width=\linewidth]{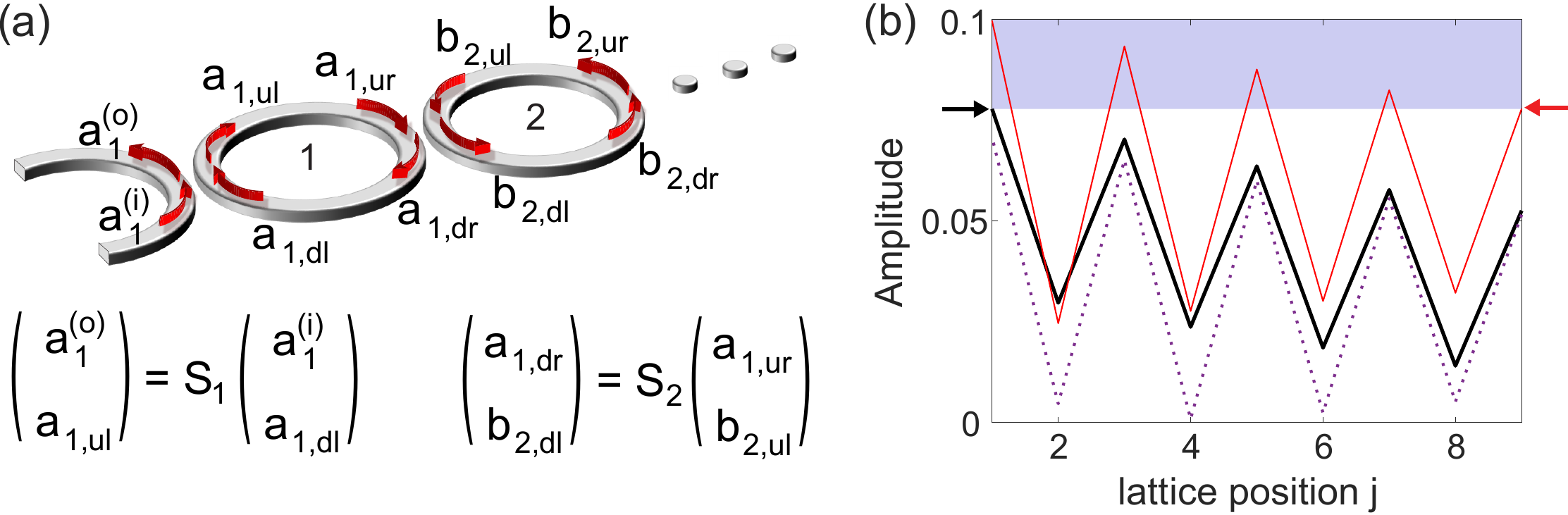}
\caption{\textbf{Transfer matrix analysis of a non-Hermitian gauged array}. (a) Amplitudes related by scattering matrices at coupling junctions. $a$'s and $b$'s are in the cavity and auxiliary rings, respectively. ``$u$'' and ``$d$'' in the subscripts indicate the upper and lower half of each ring, while ``$l$'' and ``$r$'' indicate the left and right coupling junctions. (b) Same as Fig.~\ref{fig:gauged_simulation}(e). Thick solid and dashed lines plot $a_{p,ul}$ and $a_{p,dr}$ respectively when driving from the left, and thin solid line plots $a_{p,dr}$ when driving from the right. Here $p$ counts all rings while $j$ counts only cavity rings.} \label{fig:Tmatrix}
\end{figure}

In this section, we apply and extend a transfer matrix approach \cite{Ge_PRB2023} to take into account the spatial dependency of the steady-state field inside each ring in the photonic simulations. For simplicity, we consider just the micro-ring resonators and the left and right waveguides, with the effect of radiation loss in the cavity rings and their coupling losses due to the semi-circular waveguides represented by a small positive imaginary part of the refractive index. 

For the CW modes considered in Figs.~\ref{fig:gauged_simulation}(b,c) in the main text and compared to the tight-binding model, we define their transfer matrix $M$ by
\be
\begin{pmatrix}
a^{(\text{o})}_1 \\
a^{(i)}_1
\end{pmatrix}
=
M
\begin{pmatrix}
a^{(\text{o})}_{N_a+1} \\
a^{(i)}_{N_a+1}
\end{pmatrix},\,
M \equiv
(\Pi_{p=1}^{N_a} C_pM_p)C_{N_a+1}. \label{eq:M}
\ee
Here $N_a=2N-1$ is the total number of rings (cavity rings plus auxiliary rings), $M_p\,(p=1,2,\ldots,N_a)$ is the propagation matrix within each ring, i.e.,
\be
M_p = 
\begin{pmatrix}
0 & e^{-in_{p,d}kL_{p}} \\
e^{in_{p,u}kL_{p}} & 0
\end{pmatrix},\label{eq:Mp}
\ee 
where $k$ is the free-space wave vector. $n_{p,d}$, $n_{p,u}$ are the complex refractive index along the upper and lower halves of each ring, and $L_{p}$ is their equal path length. In a cavity ring, $n_{p,u},n_{p,d}$ inside $M_p$ need to be exchanged in principle but here they are the same (i.e., uniform loss in a cavity ring). 

The coupling matrix $C_p$ is related to the scattering matrix $S_p$ at each coupling junction [see Fig.~\ref{fig:Tmatrix}(a)], i.e., 
\be
C_p =  \frac{1}{iJ_p^*}
\begin{pmatrix}
s_p & -1\\
1 & -s_p^*
\end{pmatrix},
\quad
S_p = 
\begin{pmatrix}
s_p & iJ_p\\
iJ_p^* & s_p^*
\end{pmatrix},
\ee
with $S_1$ ($S_{N_a+1}$) describing the coupling between the left (right) waveguide and the first (last) ring. $S_p$ is unitary due to flux conservation, and consequently $\det(C_p)=-J_p/J^*_p$. 

\subsection*{Non-Hermitian gauged reciprocity}

Below we first verify the non-Hermitian gauged reciprocity, given by Eq.~(\ref{eq:gauged_reciprocity}) in the main text. We note that the amplitudes of the transmission coefficients are given by $|t_L| = |a^{(\text{o})}_{N_a+1}/a^{(i)}_1|$ (with $a^{(i)}_{N_a+1}=0$) and $|t_R| = |{a^{(\text{o})}_1}/{a^{(i)}_{N_a+1}}|$ (with $a^{(i)}_1=0$) respectively, and with the extra phase introduced by the $S_p$'s removed, we find 
\be
\frac{t_R}{t_L} = -\det(M)\prod_{p=1}^{N_a+1} \frac{J_p^*}{J_p} = \prod_{j\in\text{aux}}\frac{e^{in_{p,u}kL_{j}}}{e^{in_{p,d}kL_{j}}},\label{eq:gauged_reciprocity2}
\ee
where we have used $\det{M_p}=-1$ in the cavity rings and $\det{C_p} = -J_p/J_p^*$. 

In deriving Eq.~(\ref{eq:gauged_reciprocity2}), the removal of the phases in the transmission coefficients is implemented by first considering that all $S_p$'s are off-diagonal, i.e., with $s_p=0$. In other words, light is transferred from one cavity to another at each coupling junction completely. In this case, the complex amplitude of light acquires a phase factor $iJ_p^*/|J_p|$ when propagating from left to right at junction $p$ and $iJ_p/|J_p|$ in the opposite direction. These are the factors we have removed from $t_{L,R}$ even when $s_p$ is non-zero: 
\begin{align}
t_L &= \left.\frac{a^{(\text{o})}_N}{a^{(i)}_1\Pi_{p=1}^{N_a+1} (iJ_p^*/|J_p|)}\right|_L = \frac{1}{M_{21}\Pi_{p=1}^{N_a+1} (iJ_p^*/|J_p|)},\nonumber \\
t_R &= \left.\frac{a^{(\text{o})}_1}{a^{(i)}_N\Pi_{p=1}^{N_a+1} (iJ_p/|J_p|)}\right|_R = -\frac{\det M}{M_{21}\Pi_{p=1}^{N_a+1} (iJ_p/|J_p|)}. \nonumber
\end{align}
To better understand this phase convention, we note that $t_{L,R}$ given above become $t_L=t_R=1$ as designed if we simply couple the left waveguide to the right waveguide (i.e., $N_a=0$) with total transmission (i.e., $s_1=0$ and $|J_1|=1$), which can be seen using $M=C_1$ in this case and $M_{21}=1/(iJ_1^*)$, $\det(M) = -J_1/J_1^*$.

The right hand side of Eq.~(\ref{eq:gauged_reciprocity2}) consists of only contributions from auxiliary rings, and each fraction in it, to a good approximation, is indeed the ratio of each pair of non-reciprocal couplings in the tight binding model \cite{Ge_PRB2023}, hence confirming the non-Hermitian gauged reciprocity given by Eq.~(\ref{eq:gauged_reciprocity}). In the case shown in Fig.~\ref{fig:Tmatrix}(b), this ratio of ${t_R}/{t_L}$ is found numerically by propagating the incident light first from the left and then from the right using Eq.~(\ref{eq:M}). Its value, which we find to be $1.90$, agrees exactly withe the analytical result given by Eq.~(\ref{eq:gauged_reciprocity2}). 
The equal-amplitude response due to the non-Hermitian gauged parity, evident in Fig.~\ref{fig:Tmatrix}(b), can also be proved similarly (see the next section).

When plotting Fig.~\ref{fig:Tmatrix}, we have used an effective index $\re{n}=2.48$ in all rings and $\im{n}=1.9\times10^{-4}$, $\pm8\times10^{-4}$ in the cavity and auxiliary rings, respectively. Their radii are taken to be $3.79$ and $3.84$ $\mu$m, in between the outer and inner radii from those in the simulation. Since the spacing between the input (output) waveguide and the first (last) ring is wider than that between two rings, we have used $s_1=s_{N_a}=0.99$ and $s_p=0.7\,(p\in[2,N_a-1])$. These parameters determine the field amplitudes inside the rings, as well as the transmission coefficients. To compare the former with those measured from the output of the semi-circular waveguides in the simulations, we have used a coupling efficiency of $6\%$ to these waveguides. 

With this set of parameters, we find that the other properties of the photonic system obtained from the transfer matrix analysis also closely resemble those from the full-wave simulations. For example, we mentioned the strong up-and-down amplitude asymmetry inside a subset (i.e., the even-numbered) of cavity rings in Fig.~\ref{fig:gauged_simulation} of the main text. This feature is captured well by the strong contrast of $a_{p,ul}$ and $a_{p,dr}$ in these cavities [see the thick solid and dashed lines in Fig.~\ref{fig:Tmatrix}(b)]. Note that these amplitudes change little ($\sim1\%$) when light propagates on each half of a cavity ring, due to the latter's small size and minute loss.  

\subsection*{Non-Hermitian gauged symmetry}

To prove the equal-amplitude response warranted by the non-Hermitian gauged parity symmetry in the transfer matrix analysis, we first specify the non-Hermitian gauge transformation in this framework. Because the non-reciprocal couplings are implemented using the auxiliary rings here, a gauge transformation only involves a transformation of $M_p$ defined by Eq.~(\ref{eq:Mp}), i.e.,
\be
M_p = 
\begin{pmatrix}
0 & e^{-in_{p,d}kL_{p}} \\
e^{in_{p,u}kL_{p}} & 0
\end{pmatrix}
\equiv
\begin{pmatrix}
0 & e_{p,d} \\
e_{p,u} & 0
\end{pmatrix}.
\ee 
If we rewrite $M_p$ using either off-diagonal element as a prefactor, the remaining element has an exponent proportional to $(n_{p,d}+n_{p,u})\equiv 2\bar{n}_p$. Therefore, we can construct a new system $\bar{H}$ with 
\be
\bar{M}_p = 
\begin{pmatrix}
0 & e^{i\bar{n}_{p}kL_{p}} \\
e^{i\bar{n}_{p}kL_{p}} & 0
\end{pmatrix}
\ee
that presents reciprocal couplings. The new total transfer matrix $\bar{M}$ differs from the original $M$ only by a multiplicative factor.

With this non-Hermitian gauge transformation applied to all auxiliary rings, if we require the resulting system to have parity, then we need to have
\be
\bar{M}_p = \bar{M}_{N_a-p+1}
\ee  
for all auxiliary rings and 
\be
M_p = M_{N_a-p+1}
\ee  
for all cavity rings. These two conditions can be combined to give
\be
M_p \propto M_{N_a-p+1}.\label{eq:Tmatrix_cond1}
\ee  
In addition, the parity of $\bar{H}$ also requires
\be
s_p = s_{N_a-p+2}^*,\quad J_p = J_{N_a-p+2}^*,\label{eq:Tmatrix_cond2}
\ee  
from our definition of the scattering matrix. Note that we have $N_a$ rings in total but $N_a+1$ couplings junctions, with the input and output waveguides considered.

With these observations, our proof of the equal-amplitude response warranted by the non-Hermitian gauged parity symmetry can be carried out using the original transfer matrix $M$ without applying the non-Hermitian gauge transformation. We first focus on the product of the three central matrices inside the definition of $M$, i.e., $C_{N}M_NC_{N+1}$, and we find the resulting $2\times2$ matrix is traceless:
\be
C_{N}M_NC_{N+1} \propto 
\begin{pmatrix}
-s_N^*e_{N,d}+s_Ne_{N,u} & e_{N,d}-s_N^2e_{N,u} \\
e_{N,u}-(s_N^*)^2e_{N,d} & s_N^*e_{N,d}-s_Ne_{N,u}
\end{pmatrix},\nonumber
\ee 
where we have used Eq.~(\ref{eq:Tmatrix_cond2}) with $p=N$. Next, by padding this result from the left and right by $M_{N-1}$ and $M_{N+1}$ and using Eq.~(\ref{eq:Tmatrix_cond1}) with $p=N-1$, we again find that the resulting matrix is traceless. By repeating this procedure and using Eqs.~(\ref{eq:Tmatrix_cond1}) and (\ref{eq:Tmatrix_cond2}) alternately, we find that the total transfer matrix $M$ is also traceless, i.e.,
\be
M_{11} = -M_{22}. \label{eq:Tmatrix_diagonal}
\ee  
This result is independent of whether we have an odd or even number of cavity rings.

If $\bar{H}$ has balanced gain and loss (with radiation loss also considered), then a more concise way to derive this relation is via the total scattering matrix $\bar{S}$ for $\bar{H}$, which is still unitary in this case:
\be
\begin{pmatrix}
\bar{a}^{(\text{o})}_1\\
\bar{a}^{(\text{o})}_{N_a+1}
\end{pmatrix}
=
\bar{S}
\begin{pmatrix}
\bar{a}^{(\text{i})}_1 \\
\bar{a}^{(i)}_{N_a+1}
\end{pmatrix},\quad
\bar{S}=
\begin{pmatrix}
\bar{s} & i\bar{J} \\
i\bar{J}^* & \bar{s}^*
\end{pmatrix}.
\ee
The parity of $\bar{H}$ then requires $\bar{s}=\bar{s}^*$, similar to Eq.~(\ref{eq:Tmatrix_cond2}). In other words, $\bar{s}$ is real, which in turn shows that
\be
\bar{M} =  \frac{1}{iJ^*}
\begin{pmatrix}
\bar{s} & -1\\
1 & -\bar{s}^*
\end{pmatrix}
\ee
is traceless. Now since the original $M$ and $\bar{M}$ only differ by a multiplicative factor as mentioned previously, we then recover Eq.~(\ref{eq:Tmatrix_diagonal}).

Finally, we verify that $a_{1,ul}$, when driving the lattice from the left, equals $a_{N_a,dr}$, when driving from the right, which is the manifestation of the equal-amplitude response warranted by the non-Hermitian gauged parity symmetry shown in Fig.~\ref{fig:Tmatrix}(b)]. It is easy to find 
\begin{align}
a_{1,ul} &= \frac{1}{iJ_1}\left(s_1^*\frac{M_{11}}{M_{21}}-1\right)a^{(i)}_1,\\
a_{N_a,dr} &= \frac{1}{iJ_{N_a+1}^*}\left(s_{N_a+1}\frac{-M_{22}}{M_{21}}-1\right)a^{(i)}_{N_a+1},
\end{align}
and using Eq.~(\ref{eq:Tmatrix_cond2}) again with $p=1$ and Eq.~(\ref{eq:Tmatrix_diagonal}), we finally derive
\be
a_{1,ul} = a_{N_a,dr}
\ee
with the same input amplitude in the two driving scenarios, i.e., $a^{(i)}_1=a^{(i)}_{N_a+1}$.


\end{document}